\documentclass[12pt]{article}
\newcommand{\1}{\begin{equation}}
\newcommand{\2}{\end{equation}}
\newcommand{\ea}{\begin{eqnarray}}
\newcommand{\ee}{\end{eqnarray}}
\newcommand{\bee}{\begin{eqnarray*}}
\newcommand{\eee}{\end{eqnarray*}}

\newcommand{\op}[1]{\hat{#1}}
\newcommand{\erw}[1]{\left\langle\, #1\,\right\rangle}

\newcommand{\de}{{\!\rm d}}
\newcommand{\e}{{\rm e}}

\newcommand{\g}{{\!\,=\,\!}}

\newcommand{\ii}{{\rm i}}

\newcommand{\sa}{\left[ \begin{array} {c} }
\newcommand{\se}{\end{array}\right]}

\usepackage{texdraw}
\usepackage{graphicx}
\begin{document}

\begin{center}
{\Large\bf Dynamics of Uniform Quantum Gases, II: Magnetic Susceptibility}\\*[8mm] {\large J. Bosse}\\*[1.5mm]
{\small Institute of Theoretical Physics, Freie Universit{\"a}t,
Berlin 14195, Germany}\\*[3mm] {\large K. N. Pathak}\\*[1.5mm]
{\small Department of Physics, Panjab University, Chandigarh 160
014, India}\\*[3mm] {\large G. S. Singh}\\*[1.5mm] {\small
Department of Physics, Indian Institute of Technology Roorkee, Roorkee 247
667, India\\*[3mm]
(15 December 2009)}\\
*[1cm]

\end{center}

\begin{abstract}

A general expression for temperature--dependent magnetic susceptibility of quantum gases composed of particles possessing both charge and spin degrees of freedom has been obtained within the framework of the generalized random phase approximation. The conditions for the existence of dia--, para--, and ferro--magnetism have been analyzed in terms of a parameter involving single-particle charge and spin. The limit $T\to0$ retrieves the expressions for the Landau and the Pauli susceptibilities for an electron gas. It is found for a Bose gas that on decreasing the temperature, it passes either through a diamagnetic incomplete Meissner-effect regime or through a paramagnetic--ferromagnetic large magnetization fluctuation regime before going to the Meissner phase at $T=T_{\rm BEC}$.
\\*[5mm]
\\*[5mm]

\end{abstract}

\section{Introduction}
\label{Introduction}

We wish to draw attention to an interesting application of  analytical expressions derived in Paper I \cite{bps.a:10} for the {\em particle--current} and the transverse {\em charge--current} static susceptibilities of uniform gases of {\em neutral} quantum particles. We consider here ${\cal N}$ quantum particles, each carrying a charge $(Ze)$ and a spin ${\bf s}$, moving in a volume $V$ in the presence of a neutralizing background of opposite charges and thus forming a quantum mono--plasma (``jellium''). The dimensionless temperature--dependent magnetic susceptibility of such a fluid, which may be deduced as static limit from Ref.\cite[Eq.(A16)]{raj:72}, is given by
\1
\label{def-magnetic-susc}
\chi_{\rm
m}=\lim_{H\to0}\frac{{\bf M}\cdot{\bf H}}{H^2}=\lim_{q\to0}\frac{ \omega_{\rm p}^2}{c^2q^2}
\left[\;\tilde{\chi}^C_\perp(q)/(\epsilon_0 \omega_{\rm p}^2)~-~1\;\right],
\2
where ${\bf M}$, ${\bf H}$, $\epsilon_0$, $c$, $\omega_{\rm p}$, and
$\tilde{\chi}^C_\perp(q)$ denote, respectively, magnetization, magnetic field, permittivity of vacuum, speed of light, plasma frequency, and static transverse charge--current susceptibility.
For the mono--plasma, one has $\omega_{\rm p}^2\g (Ze)^2n/(\epsilon_0 m)$ with mass $m$ and average number density $n\g {\cal N}/V$ of particles.

Choosing a cartesian coordinate system $\{{\bf e}_x,\,{\bf e}_y,\,{\bf e}_z\}$ such that ${\bf q}\g q{\bf e}_z$, the static susceptibility of the transverse charge--current density is given by
\ea
\label{stat-chi_C-T}
\tilde{\chi}^C_\perp(q)&=&\frac{\ii}{\hbar V}\lim_{\epsilon\to0}\int_0^\infty\de t~\e^{-\epsilon t}
\erw{\left[\hat{{\bf J}}_{{\bf q},\,x}^{C}(t)^\dagger,\,\hat{{\bf J}}_{{\bf q},\,x}^{C}(0)\right]}\\
\label{stat-chi_C-T-RPA}
&\approx&\epsilon_0\omega_{\rm p}^2\left\{1+\frac{\left[\tilde{\chi}^C_\perp(q)^{(0)}/(\epsilon_0\omega_{\rm p}^2)-1\right]}{1-\frac{\omega_{\rm p}^2}{c^2q^2}\left[\tilde{\chi}^C_\perp(q)^{(0)}/(\epsilon_0\omega_{\rm p}^2)-1\right]}\right\}\;,
\ee
where $\hat{{\bf J}}_{{\bf q},\,x}^C$ denotes the operator of transverse {\em charge--current} density. In the last line, $\tilde{\chi}^C_\perp(q)$ has been approximated in terms of the transverse charge--current susceptibility $\tilde{\chi}^C_\perp(q)^{(0)}$ of a fictitious similar system in which interactions between particles have been switched off \cite{pin:99}. This generalized random phase approximation (GRPA) correctly accounts for the transverse electromagnetic shielding effects in systems of particles which carry a charge and/or a spin. The conventional RPA expression for the transverse {\em particle--current} susceptibility of a system of neutral and spinless particles, Eq.(45) of Ref. \cite{bps.a:10}, is easily recovered from Eq.(\ref{stat-chi_C-T-RPA}) (taking into account Eq.(\ref{J-C-T-op}) below) for $s\g0$ and $Z\to0$.

There are three contributions to the operator of charge--current density \cite{mar:68} resulting in the transverse component
\1
\label{J-C-T-op}
\hat{{\bf J}}_{{\bf q},\,x}^C=(Ze)\hat{{\bf J}}_{{\bf q},\,x}-\ii q\gamma \hat{{\bf S}}_{{\bf q},y}-\frac{\epsilon_0\omega_{\rm p}^2}{{\cal N}}\sum_{\bf k}\left({\bf e}_x\cdot{\bf A}_{{\bf k}}\right)\delta\op{N}_{\bf q-k}
\2
with $\hat{{\bf J}}_{\bf q}$ denoting the number--current density, $\hat{{\bf S}}_{\bf q}\g\sum_{{\bf k}\sigma\sigma'}\hat{{\bf s}}_{\sigma\sigma'}a^\dagger_{{\bf k}\sigma}a_{{\bf k}+{\bf q}\sigma'}$ the spin density, $\gamma\g g\mu_{\rm B}/\hbar$ the gyromagnetic ratio with $g$ as the $g$--factor and $\mu_{\rm B}\g e\hbar/(2m_{\rm e})$ the Bohr magneton, $\op{N}_{\bf q}$ the particle--number density and its fluctuation $\delta\op{N}_{\bf q}\g\op{N}_{\bf q}-\erw{\op{N}_{\bf q}}$, and ${\bf A}_{\bf q}\g\int\de^3r~\e^{-\ii{\bf q}\cdot{\bf r}}{\bf A}({\bf r})$ the vector potential associated with the moving charges.
However, in a uniform system only the first two terms on the right--hand side will have main contributions to the transverse charge--current susceptibility in Eq.(\ref{stat-chi_C-T}), because the contribution due to the vector potential ${\bf A}_{\bf k}$ produced by the moving charges will be negligibly small in a non--relativistic gas. Moreover, cross correlations between $\hat{{\bf J}}_{{\bf q},\,x}$ and $\hat{{\bf S}}_{{\bf q},\,y}$ will also vanish due to rotational symmetry.
Therefore, we expect only two contributions to the static charge--current susceptibility: an `orbital contribution' induced by the {\em macroscopic current} density  due to moving charges and a `spin contribution' induced by the {\em magnetization--current} density due to magnetic moments associated with particle's spin.

Consequently, we express the transverse charge--current susceptibility of the noninteracting system, which is needed in Eq.(\ref{stat-chi_C-T-RPA}), as
\ea
\label{id-gas-chi_C-T}
\frac{\tilde{\chi}^C_\perp(q)^{(0)}}{\epsilon_0\omega_{\rm p}^2}&=&\frac{1}{\epsilon_0\omega_{\rm p}^2}\left[(Ze)^2\tilde{\chi}_\perp(q)+q^2\gamma^2\tilde{\chi}^S_{yy}(q)\right]\nonumber\\
&=&\tilde{\chi}_\perp(q)\frac{m}{n}~+~\frac{\gamma^2}{\epsilon_0\omega_{\rm p}^2} \,\frac{s(s+1)\hbar^2}{3}\,q^2\tilde{\chi}(q)\;.
\ee
Here $\tilde{\chi}_\perp(q)$ and $\tilde{\chi}^S_{yy}(q)$ denote the ideal--gas susceptibilities of the particle--current density and the spin density, respectively. The latter is easily shown to be proportional to the number--density susceptibility $\tilde{\chi}(q)$ of an ideal quantum gas. The expressions for ideal--gas static susceptibilities $\tilde{\chi}_\perp(q)$ and $\tilde{\chi}(q)$ have been obtained in Paper I. For   notational simplicity, superscripts $^{(0)}$ on these quantities are being omitted now onwards.

\section{Magnetic Susceptibility}
\label{Magnetic Susceptibility}

The expression for the magnetic susceptibility within GRPA is finally obtained -- using Eqs. (\ref{def-magnetic-susc}),   (\ref{stat-chi_C-T-RPA}) and (\ref{id-gas-chi_C-T}) -- in the form
\1
\label{magnetic-susc-GRPA}
\chi_{\rm
m}=\frac{\Xi(T)}{1-\Xi(T)}\;,
\2
where
\1
\label{def-Xi}
\Xi(T)=\lim_{q\to0}\frac{\omega_{\rm p}^2}{c^2q^2}\left[\tilde{\chi}_\perp(q)\frac{m}{n}-1~+~\frac{\gamma^2}{\epsilon_0\omega_{\rm p}^2} \,\frac{s(s+1)\hbar^2}{3}\,q^2\tilde{\chi}(q)\right].
\2
It is to be noted that $\Xi(T)$ may be interpreted as a static susceptibility, too, which describes the linear response of the magnetization ${\bf M}$ to the magnetic induction field ${\bf B}$. From   Eqs.(\ref{def-magnetic-susc}) and (\ref{magnetic-susc-GRPA}) in conjunction with the general relation ${\bf B}\g\mu_0(1+\chi_{\rm m}){\bf H}$, where $\mu_0\g1/(\epsilon_0c^2)$ is the free-space permeability, one finds
\1
\lim_{B\to0}\frac{{\bf M}\cdot{\bf B}}{B^2}=\Xi(T)\,\epsilon_0c^2\;.
\2
In order to evaluate $\Xi(T)$,  we read for the small--$q$ region from Eqs. (34) and (36) of Paper I:
\ea
\label{kappaofqto0}
q^2\tilde{\chi}(q)&=&\frac{4nm}{\hbar^2}\left\{C_0+\,(1-C_0)\,\frac{\beta\hbar^2q^2}{4m}\,
\frac{\zeta_{1/2}\left(\eta\lambda\right)}{\zeta_{3/2}\left(\eta\lambda\right)}+{\cal O}(q^4)\right\},
\ee
and
\ea
\label{chi-trans-asymp}
\tilde{\chi}_\perp(q)\frac{m}{n}-1&=&-\left\{C_0+(1-C_0)\,\frac{\beta\hbar^2q^2}{12m}\,
\frac{\zeta_{1/2}\left(\eta\lambda\right)}{\zeta_{3/2}\left(\eta\lambda\right)}
+{\cal O}(q^4)\right\}.
\ee
Here $\beta\g1/(k_{\rm B}T)$ and $\lambda\g\exp[\beta \mu_\eta(n,T)]$ denotes the fugacity with chemical potential $\mu_\eta(n,T)$, where $\eta\g-1,\,0$, $+1$ for the gas obeying FD, MB, BE statistics, respectively. $C_0\g \lim_{N\to\infty}N_0/N$, the fraction of particles occupying the zero--momentum single--particle state, is given by $C_0\g\delta_{\eta,\,1}\,\Theta\left(T_{\rm BEC}-T\right) \left[1-\left(T/T_{\rm BEC}\right)^{3/2}\right]$ with $\Theta(x)$ denoting the unit--step function and $T_{\rm BEC}$ the BEC critical temperature. Also, $\zeta_\nu(x)\g\sum_{i\g1}^\infty x^i/i^\nu$ denotes the polylogarithm.

Inserting Eqs.(\ref{kappaofqto0}) and (\ref{chi-trans-asymp}) into Eq.(\ref{def-Xi}), one gets
\1
\label{Xi-susc}
\Xi(T)=\left\{\begin{array}{ccc}
{\rm sign}(\xi)\,\frac{\zeta_{1/2}\left(\eta\lambda\right)}{\zeta_{3/2}\left(\eta\lambda\right)}\,\frac{T_0}{T}& {\rm if}&C_0=0\\
{\rm sign}(\xi-2Z^2)\times\infty&{\rm if}&C_0>0
\end{array}
\right.
\2
with the ``Curie constant''
\1
\label{Curie-Const}
T_0=|\xi|\,\frac{\hbar^2\,e^2n}{12\,\epsilon_0c^2\,m^2\,k_{\rm B}}
\2
and the parameter
\1
\label{xi-define}
\xi=s(s+1)g^2\left(m/m_{\rm e}\right)^2-Z^2
\2
determining whether the gas will show paramagnetic ($\xi>0$) or diamagnetic ($\xi<0$) behaviour.

Thus for neutral particles of nonzero spin ($Z\g0, ~s\ge1/2~\Longrightarrow~\xi>0$, paramagnetic) one finds for the ``Curie constant'' the well known expression
\1
T_0=\frac{\mu_0 \,s(s+1)(g\mu_{\rm B})^2n}{3k_{\rm B}}
\2
which does not depend on particle's mass $m$. The Curie law is recovered for a Boltzmann gas since the ratio $\zeta_{1/2}(\eta\lambda)/\zeta_{3/2}(\eta\lambda)$ becomes unity.   On the other hand, for spinless charged particles ($Z>0,~s\g0~\Longrightarrow~\xi<0$, diamagnetic), one finds
\1
T_0=\frac{(\hbar\omega_{\rm p})^2}{12 mc^2k_{\rm B}},
\2
which varies as $m^{-2}$. The general expressions derived by us here give the diamagnetic behaviour of spinless non--degenerate charged bosons consistent with the result obtained in Ref. \cite{hom:00}.
Furthermore, we find that a  paramagnetic gas will show a ferromagnetic ordering transition ($\chi_{\rm m}\g\infty$) at a Curie temperature $T_{\rm C}$ determined by $\Xi(T_{\rm C})\g1$ as per Eq.(\ref{magnetic-susc-GRPA}). On the other hand, a diamagnetic Bose gas will exhibit the Meissner--Ochsenfeld effect ($\chi_{\rm m}\g-1$), also called simply the Meissner effect, in the Bose-condensed phase ($0\le T\le T_{\rm BEC}$) since $|\Xi(T)|\g\infty$ from Eq.(\ref{Xi-susc}).

\begin{figure}\begin{center}
\includegraphics[width=110mm,angle=0]{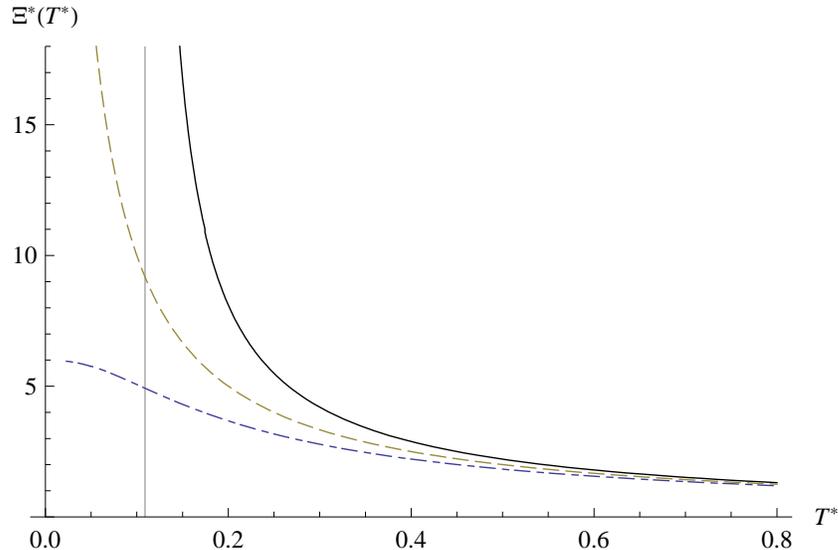}
\parbox{120mm}{\caption{\label{XiofT}\small
Temperature--dependent reduced susceptibility for BE (full line), FD (dash--dotted), and MB (dashed) gas. The vertical gridline indicates the location of $T_{\rm BEC}$ of the ideal Bose gas.  }}\end{center}
\end{figure}

In Fig.\ref{XiofT}, the (dimensionless) reduced susceptibility
\1
\label{reduced-Xi}
{\Xi}^*(T^*)\equiv\Xi(T)\frac{{\rm sign}(\xi) \epsilon_{\rm u}}{k_{\rm B}T_0}
=\frac{1}{T^*}\,\frac{\zeta_{\frac{1}{2}}\left(\eta\,\lambda^* \right)}{
\zeta_{\frac{3}{2}}\left(\eta\,\lambda^* \right)}
\2
with $\lambda^*=\e^{\mu_\eta^*(n,T^*)/T^*}$ is plotted as a function of the reduced temperature $T^*\g k_{\rm B} T/\epsilon_{\rm u}$, where $\epsilon_{\rm u}\g\hbar^2k_{\rm u}^2/(2m)$ and $k_{\rm u}\g2\left(6\pi^2n/(2s+1)\right)^{1/3}$. It can be seen that
$\Xi^*$ for a BE (an MB) gas diverges at $T\g T_{\rm BEC}$ ($T\g0$), while $\Xi^*(0)\g6$ for an FD gas.

\begin{figure}\begin{center}
\includegraphics[width=110mm,angle=0]{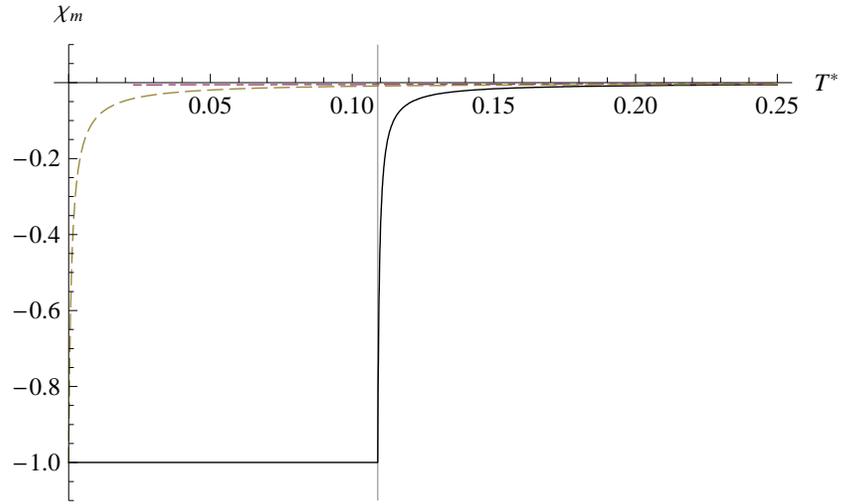}
\parbox{120mm}{\caption{\label{chimagn}\small
Temperature dependence of susceptibility for diamagnetic systems:  BE (full line), FD (dash--dotted), and MB (dashed) gas for $T_0^*\g10^{-3}$. Vertical gridline as in Fig. \ref{XiofT}.}}\end{center}
\end{figure}

\begin{figure}\begin{center}
\includegraphics[width=110mm,angle=0]{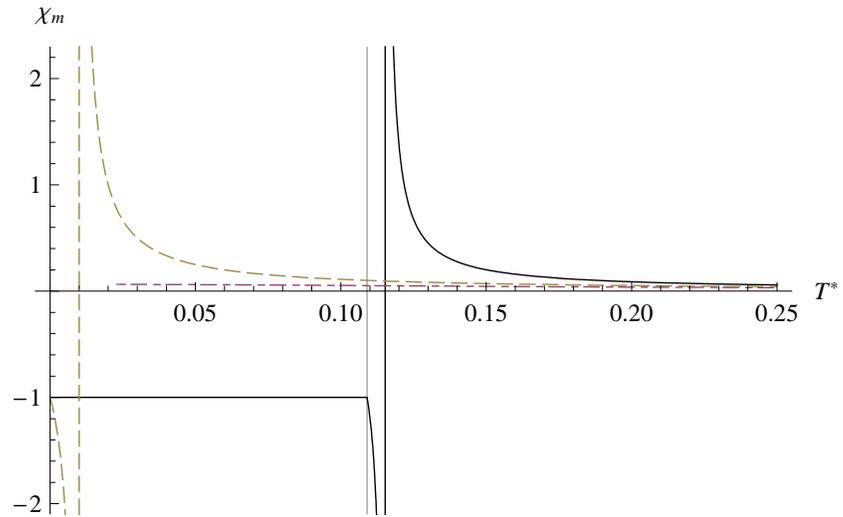}
\parbox{120mm}{\caption{\label{chiparamagn}\small
Temperature dependence of  susceptibility for paramagnetic systems: BE (full line), FD (dash--dotted), and MB  (dashed) gas for $T_0^*\g10^{-2}$. Vertical gridline as in Fig. \ref{XiofT}. }}\end{center}
\end{figure}

Typical values of $T_0^*$ are of order $10^{-6}$. Hence the magnetic susceptibility of a Fermi gas will be given, to a good approximation, by
\1
\label{Xi-approx}
\chi_{\rm m}=\frac{\Xi^*(T^*)}{\frac{{\rm sign}(\xi)}{T_0^*}-\Xi^*(T^*)}\approx{\rm sign}(\xi)\,T_0^*~\Xi^*(T^*)\;,
\2
because $1/T_0^*\gg\Xi^*(T^*)$ at all temperatures in contrast with MD and BE gases for which $\Xi^*(T^*)$ can take very large values. In order to demonstrate such a behaviour  graphically, we have plotted $\chi_{\rm m}$ as a function of temperature for the extraordinarily large value $T_0^*\g10^{-3}$ and $\xi<0$ (diamagnetic behaviour) in Fig.\ref{chimagn} for all the three gases.  Figure \ref{chiparamagn} demonstrates the behaviour of a paramagnetic gas ($\xi>0$) for the  extraordinarily large  value $T_0^*\g10^{-2}$. For this value of $T_0$, whereas the FD gas remains paramagnetic at all $T$, the BE gas and the MB gas both show a ferromagnetic phase transition at a $T_c$ which is just above $T_{\rm BEC}$ for bosons and far below for MB gases. Also, the ferromagnetic Bose gas shows large magnetization fluctuations before going to the Meissner phase.

If we consider the limiting behaviour $T\to0$ for a Fermi gas, we find from Eqs. (\ref{reduced-Xi}) and (\ref{Xi-approx}) together with Eq. (\ref{Curie-Const}) that
\1
\label{xi-FermiT=0}
\chi^{Fermi}_m(T\to0)={\rm sign}(\xi)\,\frac{3k_{\rm B}T_0}{2\,\varepsilon_{\rm F}}
\2
from which we obtain the celebrated results, see for example \cite{asm:76}, for an ideal electron gas:
\1
\chi_{_{Landau}}=-\,\frac{1}{3}\chi_{_{Pauli}}=-\,\frac{e^2k_F}{12\pi^2\epsilon_0mc^2}\,.
\2

It is pertinent to mention here that Schafroth \cite{sch:55} had established before the advent of the Bardeen-Cooper-Schrieffer \cite{bcs:57} theory of superconductivity that an ideal charged Bose gas (CBG) below   $T_{\rm BEC}$ shows the fundamental equilibrium manifestation of a superconductor, i.e. the Meissner effect. This implies that if the system is cooled from the high temperature (normal) phase, both the Bose--condensed and the superconducting phases should appear simultaneously. The concept was supported further in works \cite{sew:90,nsz:95} on the basis of gauge invariance together with the existence of the off--diagonal long--range order in systems which, in general, could be interacting. However, the study of  Koh \cite {koh:03} asserts that a CBG with short--range repulsion would show incomplete Meissner effect prior to onset of the BEC when complete Meissner effect would appear. It can be observed that our results depicted in Fig. \ref{chimagn} support such a conclusion. The behaviour in $\chi_{\rm m}$--$T$ plane at $T>T_{\rm BEC}$ seems typically akin to that for type II superconductors in $\chi_{\rm m}$--$H$ plane wherein only partial expulsion of the magnetic flux takes place (i.e. $|\chi_{\rm m}|<1$) between the critical fields $H_{C_1}$ and $H_{C_2}$. Also, behaviour depicted in Fig. \ref{chiparamagn} suggests possibility of the role of the charged bosons possessing spin degree of freedom in understanding spin-triplet superconducting transition observed in ferromagnetic compounds, see, for example \cite{hgn:07, sng:09}.

\section{Conclusion}
\label{Conclusion}

A general expression for temperature--dependent magnetic susceptibility within GRPA has been obtained for the BE, FD, and MB gases by relating it to the long-wavelength properties of $\tilde{\chi}(q)$ and $\tilde{\chi}_{\perp}(q)$ for noninteracting particles. The expression involves both charge and spin of a particle. It is shown that the
Landau (diamagnetic) and the Pauli (paramagnetic) susceptibilities are special cases ($ T\to 0$) of our more general expression. Also, the $T$--dependence of diamagnetic $\chi_{\rm m}$ for an ideal electron gas is in conformity with the numerical results obtained in \cite{ktp:02} although the comparison has not been depicted here. In summary, our work contains a unified approach for the three gases and provides a combined treatment for the study of the magnetic susceptibility for a system in which constituent particles possess both charge and spin.

\section*{Acknowledgments}

The work is partially supported by the Indo--German (DST--DFG) collaborative research program. JB and KNP gratefully acknowledge financial support from the Alexander von Humboldt Foundation.

\bibliographystyle{unsrt}

\end{document}